# Structural Alternation Correlated to the Conductivity Enhancement of PEDOT:PSS Films by Secondary Doping

K. Itoh,[1,*] Y. Kato,[1] Y. Honma,[1] H. Masunaga,[2] A. Fujiwara,[3] S. Iguchi,[1] and T. Sasaki[1]

[1]*Institute for Materials Research, Tohoku University, Sendai 980-8577, Japan*
[2]*Japan Synchrotron Radiation Research Institute (JASRI)/SPring-8, Hyogo 679-5198, Japan*
[3]*Department of Nanotechnology for Sustainable Energy, Kwansei Gakuin University, Hyogo 669-1337, Japan*
*E-mail: ito-ke592@pref.miyagi.lg.jp
*Present address: Industrial Technology Institute, Miyagi Prefectural Government, Akedohri 2-2, Izumi-ku, Sendai 981-3206, Japan

**"Secondary doping" in poly(3,4-ethylenedioxy-thiophene):poly(styrenesulfonate) (PEDOT:PSS) is quite effective and a frequently used method for the conductivity enhancement. This simple approach, the addition of co-solvents to the PEDOT:PSS solution before film fabrication, is an essential way to derive the high electrical performance of PEDOT:PSS, but the mechanism still remains unclear. In this study, nanoscale structural changes synchronized with the conductivity enhancement via secondary doping in PEDOT:PSS films were investigated. During secondary doping with ethylene glycol near the critical dopant concentration, non-crystalized PEDOT molecules uncoupled from PSS chains and then underwent nano-crystallization. These structural changes might be the key driving force for conductivity enhancement via secondary doping.**

Poly(3,4-ethylenedioxythiophene):poly(styrenesulfonate) (PEDOT:PSS) is one of the most successful polymers utilized for practical applications. In current organic electro-optical devices, PEDOT:PSS has become indispensable because of its excellent electrical properties, atmospheric stability, and flexibility as well as the ease of film fabrication owing to its good water dispersibility [1–3]. Intensive studies of conductivity enhancement have been reported thus far, with the most notable discovery being the "secondary doping" method [4]. That is, the addition of a solvent (for example, ethylene glycol (EG) or dimethyl sulfoxide (DMSO)) to the PEDOT:PSS solution is known to increase the conductivity [5–7]. This simple and convenient technique leads to dramatic increases in the conductivity from roughly 1 to nearly 1000 S/cm in thick films.

Although the secondary doping method has been used in several studies and applications, the origin of this conductivity enhancement is still debatable and remains unresolved. The key issue related to the difficulty in understanding secondary doping is the complex material structure of this system. There is a hierarchical structure with size on the order of nanometers to sub-millimeters, consisting of a sequence of monomers, poly-ion complexes of PEDOT and PSS, core-shell nanoparticles (a conducting PEDOT nanocrystal surrounded by insulating PSS molecules), and their aggregates. Thus far, structural changes caused by secondary doping have been observed on multiple length scales. X-ray scattering experiments have demonstrated that the nano-crystallization of PEDOT molecules in a PEDOT:PSS solution was assisted by the addition of polar solvents [8,9]. Additional large-scale (mesoscopic) morphological changes have been observed using scanning probe microscopy [10–13]. Aggregated domains on the order of several tens of nanometers have been clearly observed. Furthermore, the importance of decreasing the amount of insulating PSS was identified in high-concentration (>5 wt.%) DMSO-doped PEDOT:PSS films. Secondary doping was found to lead to the macroscopic phase separation of PSS on the micrometer scale [14,15]. Owing to such complicated structural changes, the crucial driving force for conductivity enhancement is still under debate. Clarifying this issue would provide not only a deep understanding of the conduction mechanism of polymers but also contribute to the optimization of the film fabrication process.

In this study, we investigated the structural changes of PEDOT:PSS films caused by secondary doping with EG by means of grazing incidence wide-angle X-ray scattering (GIWAXS) using synchrotron radiation. This approach enabled us to clarify structural changes on the nanometer-scale, which correspond to the nano-crystallization of PEDOT molecules and/or the assembly of PEDOT and PSS. The results revealed that the conformation changes of the PEDOT fragments occur simultaneously with the conductivity enhancement. Around the critical concentration of EG, PEDOT fragments are uncoupled from PSS chains, and then assembled into PEDOT nanocrystals.

A PEDOT:PSS (Clevios™ PH 1000) stock solution was used for preparing drop-cast films. EG was used as the secondary dopant. The PEDOT:PSS solution with a specific amount of EG (0–9 wt.%) was dropped on a silicon substrate (20 mm × 20 mm × 0.525 mm); then, the substrate was dried in an electric furnace at 60 °C for 120 min and at 160 °C for 30 min. The drying condition is selected to implement that drying will be completed within a reasonable time and without boiling. To evaporate two solvents with different boiling points (water and EG), we intend to evaporate water in the first process at 60 °C and EG (the boiling point is 198 °C) in the second process at 160 °C. GIWAXS measurements were performed at BL40B2 of SPring-8 in Hyogo, Japan, at an incident X-ray wavelength and angle of 1 Å and 0.2°, respectively. The scattering profiles parallel and perpendicular to the film surface, obtained from vertical and horizontal line-cuts of the GIWAXS image, are referred to as the out-of-plane ($q_z$) and in-plane ($q_y$) profiles, respectively.

Figure 1a shows the room-temperature electrical conductivity ($\sigma_{RT}$) of the films as a function of the EG concentration. In previous studies, it was observed that $\sigma_{RT}$ drastically increased by two orders of magnitude at EG concentrations of up to 2 wt.% [5,16]. Hereafter, we define this value as the critical concentration $x_c$. It is noted that this definition is different from that used in previous studies [5,16], where the critical concentration was defined as the midpoint of the plot of $\log(\sigma_{RT})$. In contrast, the present definition corresponds to a concentration indicating saturation of the conductivity enhancement.

To elucidate the correlation between the conductivity enhancement and structural changes, GIWAXS measurements were performed. Figure 1b shows 2D GIWAXS images of films with EG concentrations of 0 and 9 wt.%. Changes in the scattering images are observed, especially for the scatterings at $q_z < 10$ nm$^{-1}$ ($q_y = 0$) and at around $(q_y, q_z) = (0, 18$ nm$^{-1})$. Figures 1c and 1d show the out-of-plane (along $q_z$) and in-plane profiles (along $q_y$), respectively. In the out-of-plane profiles, characteristic peak structures were observed at 2.1, 4.8, 6.8, 9.7, 12.5, and 18.3 nm$^{-1}$. In contrast, the in-plane profiles were featureless below 10 nm$^{-1}$ and had an additional peak at 25 nm$^{-1}$. It is noted that each position was determined from the profile of the film with 9 wt.% EG.

First, we focus on the region of $q < 10$ nm$^{-1}$ along the out-of-plane direction. To assist the understanding of the following discussion, we present in advance the resulting cross-sectional view of the schematic film structure in Figure 2a. The strong peak at 2.1 nm$^{-1}$, as shown in the left panel of Figure 2b, corresponds to the lamellar stacking of the PEDOT-rich regions and intertwined PSS molecules [6,17] with a period of 3 nm. As shown in Figure 2c, the integrated intensity of this peak gradually increased with increasing EG concentration.

The 4.8-nm$^{-1}$ hump and 9.7-nm$^{-1}$ peak correspond to the alternating stacking of PEDOT and PSS ($d = 1.3$ nm) and its secondary peak, respectively [2,18]. Notably, with regard to the $q$ values, these peak positions correspond to the expected positions of the second- and fourth-order peaks of the intense 2.1-nm$^{-1}$ peak; however, this assignment is not appropriate because of the opposite trend of the peak intensity changing with EG concentration. The strengths of both the 4.8-nm$^{-1}$ hump (middle panel of Figure 2b) and the 9.7-nm$^{-1}$ peak (right panel of Figure 2b) decreased with increasing EG concentration in contrast to the increasing behavior of the 2.1-nm$^{-1}$ peak (left panel of Figure 2b). In particular, the area of the 4.8-nm$^{-1}$ hump shows characteristic behavior at $x_c$. As shown in Figure 2d, the strength of the hump area rapidly decreased with increasing EG concentration; it reached almost zero at $x_c$ (the integrated area is specified in Figure S1). This anomalous behavior stands in contrast to the gradual increase of the 2.1-nm$^{-1}$ peak with increasing EG concentration, as shown in Figure 2c.

The weak peak observed at 6.8 nm$^{-1}$ ($d = 0.92$ nm) shown in the right panel of Figure 2b might relate to the lamellar structure because it shows a tendency similar to that of the 2.1-nm$^{-1}$ peak. The edge-to-edge stacking of PEDOT molecules is the most reasonable identification for this peak. The same peak assignment has been reported for crystalline PEDOT with a somewhat longer $d$ value of 1.05 nm [19]. Unfortunately, this peak is too weak to separate from other scatterings, thus we cannot discuss the changes in the edge-to-edge stacking. This weak intensity is in common with the previous study [9]. Considering the previous study that the edge-to-edge stacking is improved by the acid-treatment (the removing of PSSs) [20], future experiments on such films might shed light on the changes in the edge-to-edge stackings. In contrast, the changes in the face-to-face stacking can be discussed because its peak ($q = 18.3$ nm$^{-1}$) is strong enough and separable from other scatterings. The peak analysis and discussion of it will be mentioned in the latter part.

The above results based on the scatterings of $q < 10$ nm$^{-1}$ elucidate two phenomena: the gradual growth of the lamellar structure, and the uncoupling of the alternate stacking with the addition of EG. Importantly, the latter change, i.e., the process of uncoupling the alternating stacks, was observed only at low EG concentrations of up to $x_c$. This uncoupling might be originating from the high dielectric constant of EG, because it could shield the strong Coulomb interaction between charged PEDOT and PSS [5].

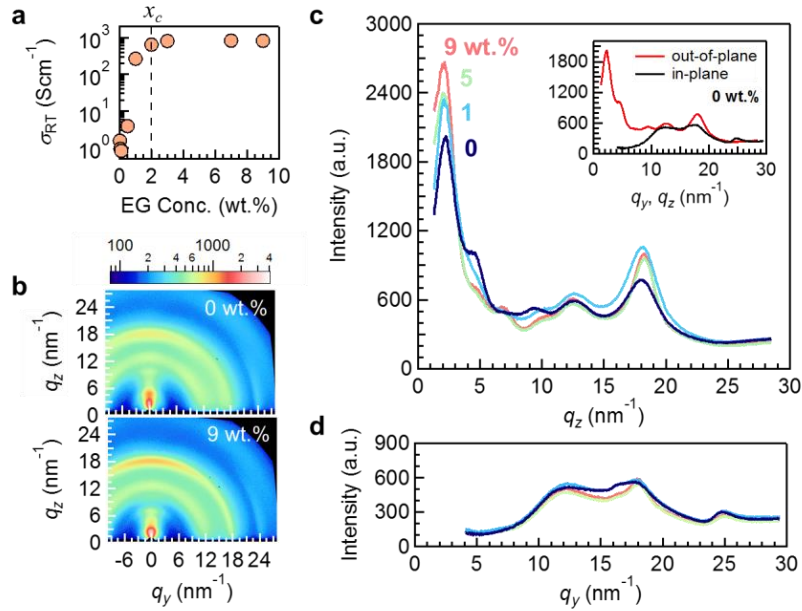

Figure 1. (a) Electrical conductivity at room temperature ($\sigma_{RT}$) as a function of the EG concentration. (b) GIWAXS profiles of 0 and 9 wt.% EG-doped films. (c), (d) Scattering profiles obtained from GIWAXS images along the out-of-plane and in-plane directions. The inset of (c) shows both the out-of-plane and in-plane profiles at 0 wt.% EG concentration.

Next, we discuss wide-angle scattering in the region of $q > 10$ nm$^{-1}$. In this region, there were large peaks in both the out-of-plane and in-plane directions, as shown in Figures 1c and 1d, respectively. The broad peak centered at 12.5 nm$^{-1}$ has been assigned to the PSS amorphous halo [21]. The major peak observed at 18.3 nm$^{-1}$ corresponds to the π–π stacking of the crystallized PEDOT [8,9]. We define this as the 010 peak [18,22] on the basis of the monoclinic unit cell, although this peak has been referred to as the 020 peak by assuming the orthorhombic unit cell in some previous studies [8,23]. In addition to these peaks, two small peaks at 16.5 and 25 nm$^{-1}$ were observed only in the in-plane profiles. These peaks can be associated with the monomer repetition in polymer chains of PEDOT and PSS [24], respectively. Hereafter, we refer to the former peak at 16.5 nm$^{-1}$ as the 001 peak.

To extract detailed information on PEDOT nanocrystallization, peak fitting to the 010 peak was performed. The quasi-Voigt function [25] with a linear background was used for peak fitting as follows:

$$P_0(q) = \eta \frac{2F_0}{\pi w_0}\left\{1 + 4\left(\frac{q-q_0}{w_0}\right)^2\right\}^{-1} + (1-\eta)\frac{2F_0}{w_0}\sqrt{\frac{\ln 2}{\pi}}\exp\left\{-4\ln 2\left(\frac{q-q_0}{w_0}\right)^2\right\} + A + Bq,$$

where $F_0$ is the peak strength, $\eta$ is the shape parameter, $q_0$ is the peak position, and $w_0$ is the full-width at half-maximum (FWHM) of the 010 peak. $A$ and $B$ are the constant and coefficient values of the linear background curve, respectively. To reproduce the overlapping of the 001 and 010 peaks, a Lorentzian function was added to the above fitting function only for in-plane analysis:

$$P_1(q) = P_0(q) + \frac{2F_1}{\pi w_1}\left\{1 + 4\left(\frac{q-q_1}{w_1}\right)^2\right\}^{-1},$$

where $F_1$, $q_1$, and $w_1$ represent the peak strength, position, and FWHM of the 001 peak, respectively.

The experimental scattering profiles and fitting curves are shown in Figure 3a (left panel: in-plane profiles, right panel: out-of-plane profiles). Good agreement was seen between the fitting curves (red) and experimental profiles (black). Figures 3b–e show the EG concentration dependence of $w_0$, $q_0$, $F_0$, and $\eta$, respectively.

As shown in the inset of Figure 3b, the FWHM showed a clear plateau up to an EG concentration of 1 wt.% and then decreased with increasing EG concentration, reflecting the intra-nanocrystal structural refinement. Notably, this decrease was significant below $x_c$ and then became moderate up to 9 wt.%. The crystallite size estimation from the FWHM is described in the Supporting Information (Figure S3). The EG dependence of the peak positions shown in Figure 3c exhibited a trend similar to the change in the peak width; the peak positions increased at 1–3 wt.% and then almost saturated. The π-π stacking distance of the face-on nanocrystal, which obtained from $q_0$ in $q_z$ profiles, at pristine and 3 wt.% are 0.348 nm and 0.343 nm, respectively. This shrinkage induced by the EG addition has good agreement with the previous study [9], that is probably because of the improvement of the crystallinity. The π-π stacking distance of the edge-on nanocrystal, which obtained from $q_y$ profiles, also showed shrinkage from 0.351 nm (pristine) to 0.347 nm (3 wt.%). Notably, there was a small difference between the π–π stacking distance of face-on nanocrystals and that of edge-on nanocrystals. The distance in the face-on nanocrystals is slightly shorter (~0.003 nm) than that in the edge-on nanocrystals at all EG concentrations. This result has good agreement with the previous study [9]. Notably, there was a small difference in the peak position between in-plane and out-of-plane directions, possibly reflecting slight differences in the π–π stacking distance between face-on and edge-on nanocrystals [9]. The peak strength shown in Figure 3d exhibited completely different EG concentration dependence from that of $w_0$ and $q_0$. In the out-of-plane profiles, the strength rapidly increased below $x_c$ and then gradually decreased, whereas almost no EG concentration dependence was observed in the in-plane profiles. In the out-of-plane direction, the values observed at 1 wt.% were 1.6 times larger than those at 0 wt.%. Considering the flat FWHM below 1 wt.%, this increase might indicate an increase in the number of a face-on nanocrystals. The shape parameter shown in Figure 3e has no characteristic EG concentration dependence. The other parameters are shown in the Supporting Information.

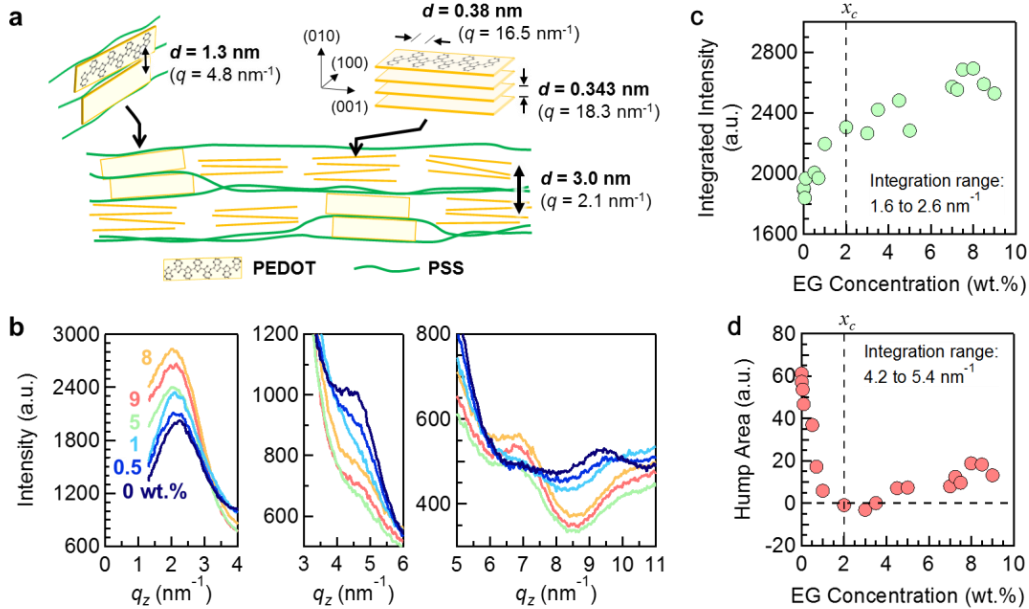

Figure 2. (a) Schematic illustration of the cross-sectional film structure. Note that the coordinate axes shown near the PEDOT nanocrystal are valid only for that nanocrystal. (b) Magnified images of the out-of-plane profiles at 0–4 (left panel), 3–6 (middle panel), and 5–11 nm$^{-1}$ (right panel). (c) Integrated intensity of the out-of-plane profiles (ranging from 1.6 to 2.6 nm$^{-1}$) and (d) area of the 4.8-nm$^{-1}$ hump as a function of the EG concentration. The hump area was estimated by the integration of the profiles between 4.2 and 5.4 nm$^{-1}$ after subtracting a linear background. Profiles to clarify the hump area are displayed in Supporting Information (Figure S1).

The structural changes caused by secondary doping are summarized as follows. At low concentrations below $x_c$, significant peeling of isolated PEDOT molecules from PSS chains occurs. Then, the uncoupling of the alternating stacks of PEDOT and PSS is expected. Simultaneously, the total volume fraction of PEDOT nanocrystals increases because the PEDOT fragments are assembled into face-on oriented PEDOT nanocrystals. Subsequently, the structural refinement of each nanocrystal occurs. This refinement, as evidenced by the decrease in the FWHM, was prominently observed around $x_c$. Importantly, as shown in Figure 1a and in previous reports [5,16], the electrical conductivity drastically increases at ~1 wt.%. This concentration dependence is quite synchronized to that of the uncoupling of the alternating stacks and the increase in the total volume fraction of nanocrystals. Therefore, it is strongly suggested that the increase in the electrical conductivity is strongly correlated with these structural changes.

We discuss this correlation from the viewpoint of inter-grain transport in polymers. As indicated in previous studies [24,26,27], the intra-lamella conduction path among PEDOT nanocrystals is important for the high electrical conductivity of PEDOT:PSS films. Thus, if isolated PEDOT molecules exist between the nanocrystals, as shown in the upper panel of Figure 4, they act as a bottleneck in the conduction path. This might be an origin of the low electrical conductivity of the pristine (0 wt.%) films. After these PEDOT fragments were uncoupled from PSS chains and assembled into nanocrystals, such bottlenecks would be removed (the lower panel of Figure 4).

It should be mentioned that while the conductivity enhancement showed almost saturation behavior around $x_c$, the peak width continuously decreased up to 9 wt.% (Figure 3b). This result suggests that the quality (crystallite size) of the nanocrystals is not predominantly related to the drastic enhancement of electrical conductivity up to $x_c$. Of course, the crystallinity of each nanocrystal should contribute to electric conductivity after removing the conduction bottlenecks among nanocrystals; thus, there is a possibility of enhancing the highest attainable electrical conductivity by improving the crystallinity. In the future, when trying to improve the conductivity in accordance with this possibility, the improvement of crystallinity with macroscopic homogeneity at high additive concentrations [12,14][12,14] should be the most important issue.

In conclusion, we elucidated the transformation of the nano-scale structure simultaneously occurring with the increase of electrical conductivity upon the secondary doping of PEDOT:PSS with EG. PEDOT fragments, which formed alternating stacks with PSS, were uncoupled from PSS and assembled into nanocrystals, at which point the conductivity drastically increased. The crystallinity of the nanocrystals was continuously improved with increasing EG concentration and had no distinct change at $x_c$. These results strongly suggest that the dramatic increases in the electrical conductivity at $x_c$ is originating from the eliminating the bottleneck in the conduction path on the nanometer scale.

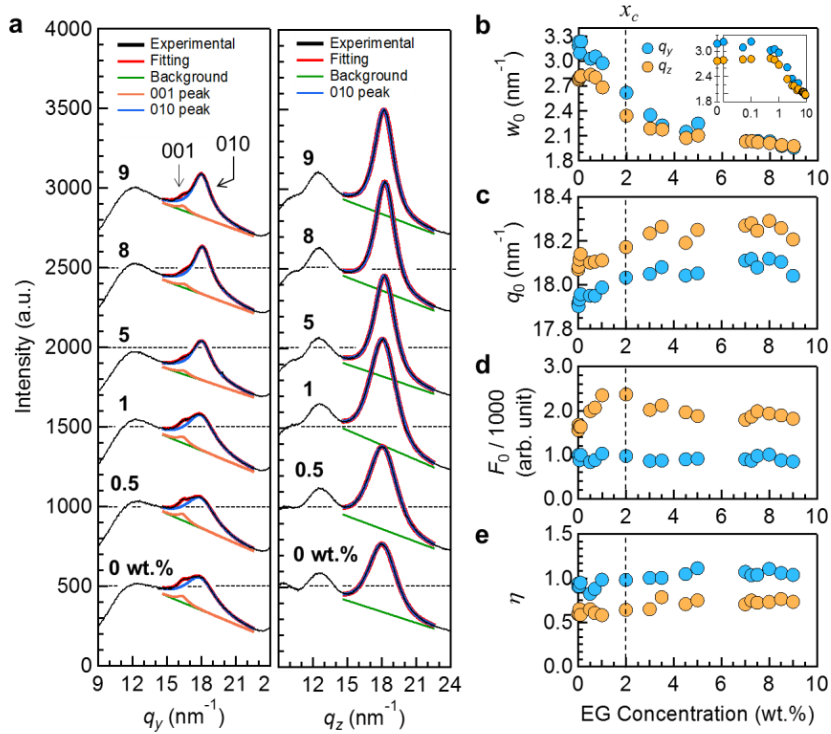

Figure 3. (a) Experimental profiles (black) and fitting curves (red) in the in-plane (left panel) and out-of-plane (right panel) directions. (b)-(e) EG concentration dependence of fitting parameters to the 010 peak: full-width at half-maximum $w_0$, position $q_0$, strength $F_0$, and shape parameter $\eta$, respectively.

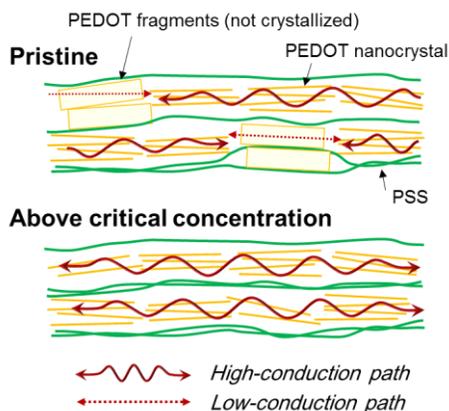

Figure 4. Schematic illustration of the structural transformation caused by secondary doping.

**Acknowledgements**

We thank H. Sekiguchi and N. Ohta for technical assistance in these experiments. We also thank T. Nishizaki, N. Asano, R. Kobayashi, and K. Hashimoto for the fruitful discussions. The synchrotron radiation experiments were performed at BL40B2 of SPring-8 with the approval of JASRI (proposal nos. 2015B1694, 2016A1617, 2017B1190). This work was supported by JSPS KAKENHI grant numbers JP15K17688 and 16K05430 and the Foundation for Promotion of Material Science and Technology of Japan.

# Supporting Information

**Structural Alternation Correlated to the Conductivity Enhancement of PEDOT:PSS Films by Secondary Doping**

Keisuke Itoh*, Yoshihisa Kato, Yuta Honma, Hiroyasu Masunaga, Akihiko Fujiwara, Satoshi Iguchi, and Takahiko Sasaki

## I. INTEGRATION OF THE HUMP STRUCTURE AT 4.8-nm$^{-1}$

To unify the integration range in all profiles, we chose the integration range of the 4.8-nm$^{-1}$ hump from 4.2 to 5.4 nm$^{-1}$, as shown in Figure S1. Black broken lines indicate the linear background and blue hatches indicate the integrated region used to obtain the hump area plotted in Figure 2d. We should mention that the hump area at low EG concentrations below 1 wt.% is underestimated. For example, in the profile at 0 wt.% (dark blue curve), the bottom edge of the hump seems to exceed the range of 4.2–5.4 nm$^{-1}$; the red hatched region was omitted. Considering this point, the decreasing of the hump area at ~1 wt.% might be more remarkable than that shown in Figure 2d.

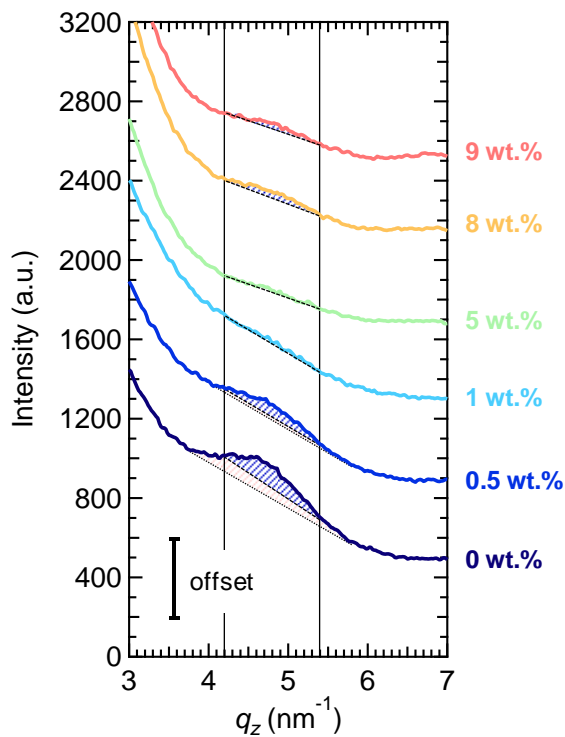

**Figure S1.** Magnified images of out-of-plane profiles at 3–7 nm$^{-1}$. Profiles are offset vertically for clarity.

## II. FITTING PARAMETERS (010 and 001 PEAKS)

Figure S1 shows the reduced activation energy $W = d(\ln \sigma_0)/d(\ln T)$ as a function of the temperature (a so-called Zabrodskii plot), which is known to be useful for differentiating among the "metallic," "critical," and "insulating" regimes in the context of the metal–insulator transition in conducting polymers.[S1,S2] It can be seen that our film was in the "metallic" regime because $W$ is not constant, but rather increases with increasing temperature in the low-temperature region.

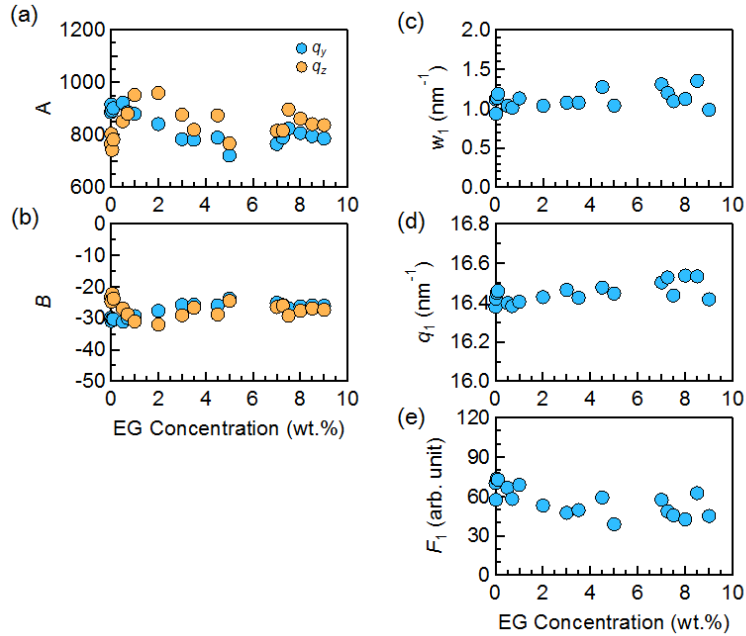

**Figure S1.** Fitting parameters for the 010 peak. (a) Constant and (b) coefficient values of the linear background. Fitting parameters for the 001 peak including the (c) FWHM, (d) position, and (e) strength. Note that the 001 peak was considered only for in-plane analysis as described in the manuscript.

## III. CRYSTALLITE SIZE ESTIMATION

The crystallite size of a PEDOT nanocrystal can be estimated from 010 peak analysis and the Scherrer equation:

$$D = \frac{K\lambda}{\beta \cos \theta}$$

where $D$ is the crystallite size, $K$ is the Scherrer constant, $\lambda$ is the X-ray wavelength, $\beta$ is the full-width at half-maximum, and $\theta$ is the Bragg angle. Figure S3 shows the calculated crystallite size using a Scherrer constant of $K = 0.9$. The crystallite sizes in $q_z$ (reflecting the face-on nanocrystal) at 0 and 9 wt.% are 2.04 and 2.86 nm, respectively. The difference between these is 0.82 nm, which is roughly twice as large as the π-π stacking distance of 0.343 nm. Crystallite size seems to have no distinct change at $x_c$. It gradually increased with increasing EG concentration. This result might suggest that the crystallinity of PEDOT is not essential to the dramatic increase of the electrical conductivity at $x_c$. It is noted that this does not indicate the crystallinity has no relation to the electrical conductivity; the highest attainable conductivity should be related to the high crystallinity. Our result indicates that the structural alternation as described in text should be necessary for the dramatic increases in the electrical conductivity at $x_c$, in addition to the continuous improvement of the crystallinity.

We should mention that the ambiguity of the peak width due to the experimental conditions (for example, the long footprint) and its effect on the Scherrer constant were not considered. Thus, the crystallite size discussed in this section is a rough estimation.

As shown in Figure S2a, the temperature dependence of the electrical conductivity in the pristine film was well represented by the exponential function ($\sigma(T) \propto \exp\left\{-\left(\frac{T_0}{T}\right)^\gamma\right\}$, where $T_0$ is the characteristic temperature and $\gamma$ is the parameter reflecting the hopping process and dimensionality). Such exponential-type temperature dependence and the value of $\gamma \sim 0.5$ indicates that the charge transport in the pristine film can be understood by the Efros-Shklovski variable-range hopping (VRH).[S3] In contrast, the EG-doped film does not show exponential-type dependence. As shown in Figure S2b, the temperature dependence of the electrical conductivity followed the power-law type dependence ($\sigma(T) \propto T^{1/2}$) at low temperatures.[S2,S4] Such changes in the temperature dependence, the exponential-type to the power-law type, indicates that tunneling process is more significant than the hopping process in the EG-doped film.

We should note that the power-law of $T^{1/2}$ in Figure S2b corresponds to the 3D WL. An ideal 2D system should follow the logarithmic dependence of $\sigma(T) \propto \log T$.[S4] This discrepancy of the dimensionality between the MC and bulk charge transport might indicate that the 2D transport is limited within the mesoscopic structure.

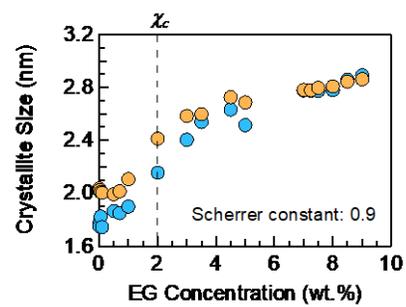

**Figure S3.** Crystallite size calculated with the Scherrer equation.